\documentclass[aps,showpacs,showkeys,letterpaper]{revtex4}
\usepackage{graphicx}
\usepackage{amsmath}
\begin{normalsize}\end{normalsize}

\begin{document}
\title{Quantum circuits under magnetic field}
\author{Constantino A. Utreras D\'{\i}az }
\affiliation{Instituto de F\'{\i}sica,
Facultad de Ciencias, Universidad Austral de Chile,
Campus Isla Teja s/n, Casilla 567, Valdivia, Chile}
\email{cutreras@uach.cl}
\author{David Laroze }

\affiliation{Instituto de Alta Investigaci\'on, Universidad de
Tarapac\'a,  Casilla 7D, Arica, Chile}
\date{\today, Valdivia}
\begin{abstract}
We consider a quantum $LC$ circuit under a constant magnetic flux $f$, and derive a discretized form of the Schr\"odinger equation, which is equivalent to introducing a {\em potential} $V(\phi,f)$ in the pseudo-flux $\phi$-representation, which is different from that found previously by Li and Chen~\cite{LI-CHEN}. Firts, we discuss the physical assumptions leading to these different results, and then study the energy spectrum of the $LC$ quantum circuit as a function of a constant external magnetic flux, using a direct numerical approach. The results are compared with the spectrum obtained using the Li-Chen potential. Our results indicate that the energy spectra from both models are quite different numerically, and as a function of $f$, so that they may be clearly distinguished under appropiate experimental conditions.
\end{abstract}
\maketitle
\section{Introduction}

In a series of articles Li and Chen~\cite{LI-CHEN,YOU-LI}  and us~\cite{FLORES-UTRERAS,FLORES-UTRERAS2,UTRERAS-FLORES,FLORES-BOLOGNA}, have developed a theory of quantum electrical systems, based on treating such systems as quantum $LC$ circuits with discrete charge. In this approach, mesoscopic and nanoscale electrical systems are described by two phenomenological parameters: an inductance $L$, and a capacitance $C$. This approach is expected to apply when the transport dimension becomes comparable with the charge carrier coherence length. In this way, one takes into account both the quantum mechanical properties of the electron system, and also the discrete nature of electric charge.

The problem of the quantum $LC$ circuit under a constant magnetic has been discussed by Li and Chen~\cite{LI-CHEN}, based on the assumption of gauge invariance of the Hamiltonian. They start from {\em field-free} discretized quantum Hamiltonian, and then, using a gauge transformation, they reintroduce the magnetic field (rather, the flux) into the equation. The crucial assumption made here is that one first finds the discrete-charge quantum Hamiltonian, and then proceeds to reintroduce the magnetic field, as described above.

In this work, we present an alternative treatment, in which we introduce charge discretization into a quantum Hamiltonian {\em with magnetic field}. The two procedures are not equivalent, and they give rise to two quite different Hamiltonians. At the present time we find no way to decide a priori which Hamiltonian is the correct one, and, perhaps, it may even be true that, under appropriate circumstances, each may be able to explain a different set of experimental situations.

This work is organized as follows: In section ~\ref{sec-quant}, we discuss the quantization process, the inclusion of magnetic field, and the subsequent charge discretization process. In Section~\ref{sec-numer} we discuss our numerical results.

\section{Quantization and Magnetic field}
\label{sec-quant} Let us review first the quantization and  charge
discretization procedure for a circuit under an external magnetic
field. Therefore, we consider a classical $LC$  circuit enclosing
an external flux $f$, with its corresponding classical Hamiltonian
${\cal H}$ is given by,
\begin{equation}
{\cal H}  =  \frac{ ( \phi - f)^2 }{2 L}  + \frac{q^2}{2 C}.
\end{equation}

In this Hamiltonian, the charge variable $q$ is similar to the
position coordinate ($x$), while the magnetic flux $\phi$ is
analogous to the momentum $p$ of the harmonic oscillator;
therefore, it may be quantized according to the usual rules,
namely, $q$ and $\phi$ become the  operators $\hat q$ and $\hat
\phi$, which satisfy the canonical commutation rules $[\hat q,\hat
\phi] = i \hbar$. As it is usual in quantum mechanics, we start
considering the wave function $ \Psi(q)$ in the charge
(coordinate) representation, in which $\hat q $ is diagonal, $\hat
q = q$ and  $\hat p  =  - i \hbar \frac{d}{dq}$. Similarly, the
Hamiltonian function ${\cal H}$ becomes the Hamiltonian operator $
\hat {\cal H}$,
\begin{equation}
\hat {\cal H} = \frac{1}{2 L} \left(    - i \hbar \frac{d}{dq} - f \right)^2 + \frac{q^2}{2 C },
\end{equation}
and the stationary Schr\"odinger equation  $\hat {\cal H} \Psi = E \Psi $ becomes
\begin{equation}
\frac{1}{2 L} \left(    - i \hbar \frac{d}{dq} - f \right)^2 \Psi(q) + \frac{q^2}{2 C } \Psi(q) = E \Psi(q).
\end{equation}
The quantization procedure described above rests on the assumption that the electric charge is actually a continuous variable, but the electric charge is known to be quantized; therefore~\cite{LI-CHEN}, one replaces the differential operator $d/dq$ by the centered difference operator
\begin{equation}
D \Psi(q) = \frac{  \Psi(q + \Delta q/2 ) - \Psi(q - \Delta q/2 )  }{\Delta q},
\end{equation}
in which $\Delta q = q_e$ may be identified with the electron charge. The Schr\"odinger equation with discrete charge, in the charge representation, becomes

\begin{equation}
\frac{1}{2 L} \left( - i \hbar D  - f \right)^2 \Psi(q) + \frac{q^2}{2 C } \Psi(q) = E \Psi(q).
\end{equation}
Notice that now we only want to find the wavefunction at the discrete points, $\Psi(q_n)$ ($q_n = n q_e$). The pseudo-flux representation is obtained by a discrete Fourier transformation, applied to the wavefunction on the charge representation, $\Psi(q_n) $,  as in Equation (\ref{Eq:pseudoflux}) below, which shows that the wavefunction in pseudo flux representation is stricly periodic in $\phi$, with period $2\pi \hbar/q_e$,

\begin{equation}
\label{Eq:pseudoflux}
\Psi(\phi) = \sum_n e^{-i (n q_e) \phi /\hbar} \Psi(q_n).
\end{equation}
In what follows, the variable $\phi$ is called the {\em pseudo flux representation}, since it has the physical dimension of a magnetic flux, and it becomes the true flux in the $ q_e \to 0$ limit. In this representation the operators become
\begin{eqnarray}
-i \hbar D &=&  \frac{2 \hbar }{q_e} \sin(\frac{q_e\phi}{2 \hbar}) \\
\hat q &=& i \hbar \frac{d}{d\phi}.
\end{eqnarray}
The discretized quantum Hamiltonian becomes
\begin{equation*}
\hat H = \frac{1}{2 L} \left( \frac{2 \hbar }{q_e} \sin(q_e \phi/2
\hbar)- f \right)^2 + \frac{\hat q^2}{2 C}.
\end{equation*}
Let us define $\phi_0 = \hbar/q_e$ (a flux quantum), and the {\em potential} $V(\phi)$,
\begin{equation}
\label{V-new}
V(\phi)  =  \frac{1}{2 L} \left( 2 \phi_0 \sin( \frac{\phi}{2\phi_0} ) - f \right)^2.
\end{equation}
This potential has a very interesting structure, as shown in Figure~\ref{Fig-pot} below. For zero external magnetic flux $f$, the period is $2\pi \phi_0$, but, once a nonzero magnetic field is connected, the period becomes $4 \pi \phi_0$. Also, the strength of the potential increases with increasing $f$, and, for intermediate values of $f$, it shows a small maximum in each period. On the other hand, Li and Chen~\cite{LI-CHEN} have proposed that the potential should be given by

\begin{equation}
\label{V-LiChen} V_{LC}(\phi)  =  \frac{1}{2 L} \left[ 2\phi_0
\sin \left( \frac{\phi}{2\phi_0} - f \right) \right]^2
\end{equation}

\begin{figure}[h]
\includegraphics[width=8.0cm]{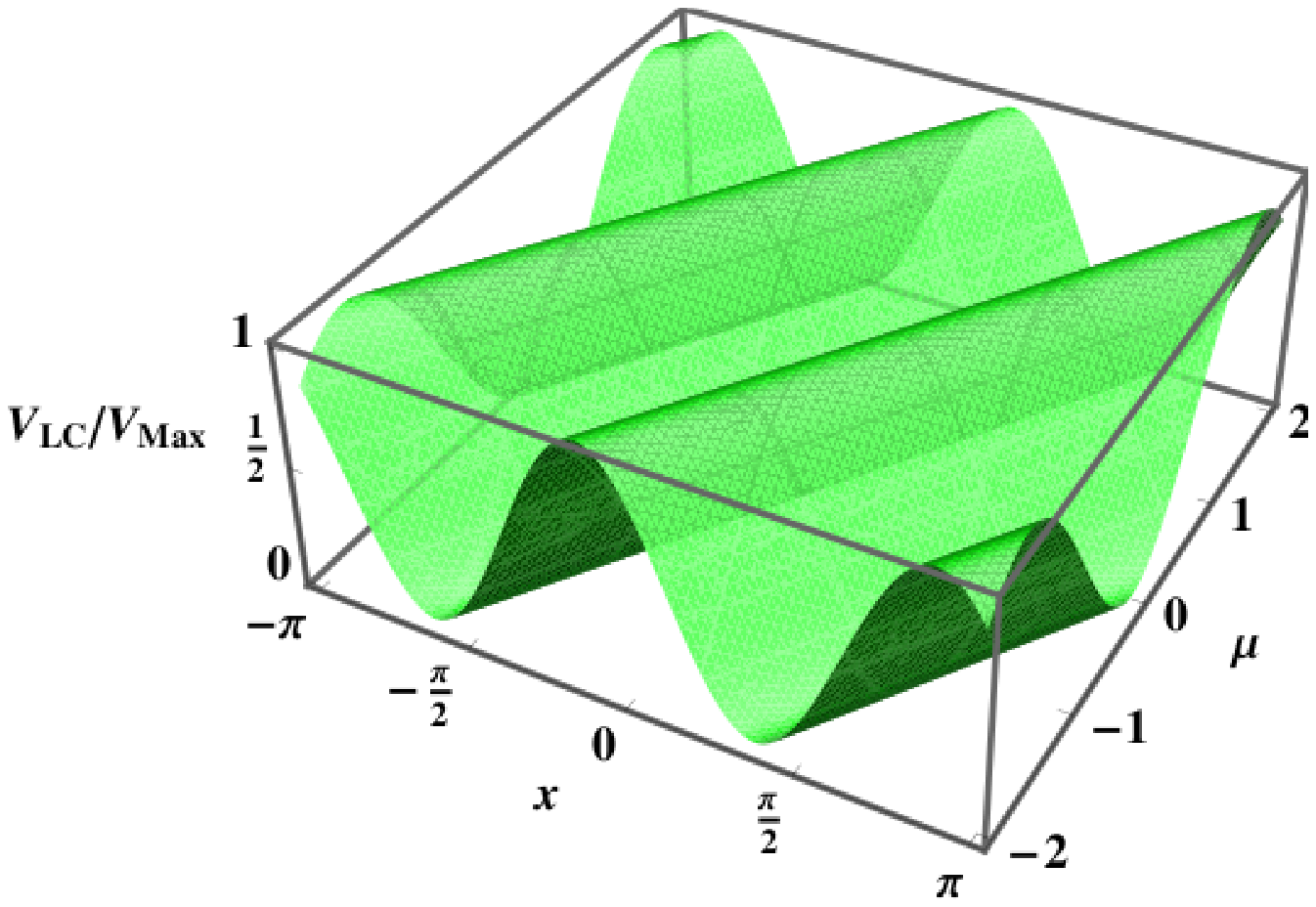}%
\includegraphics[width=8.0cm]{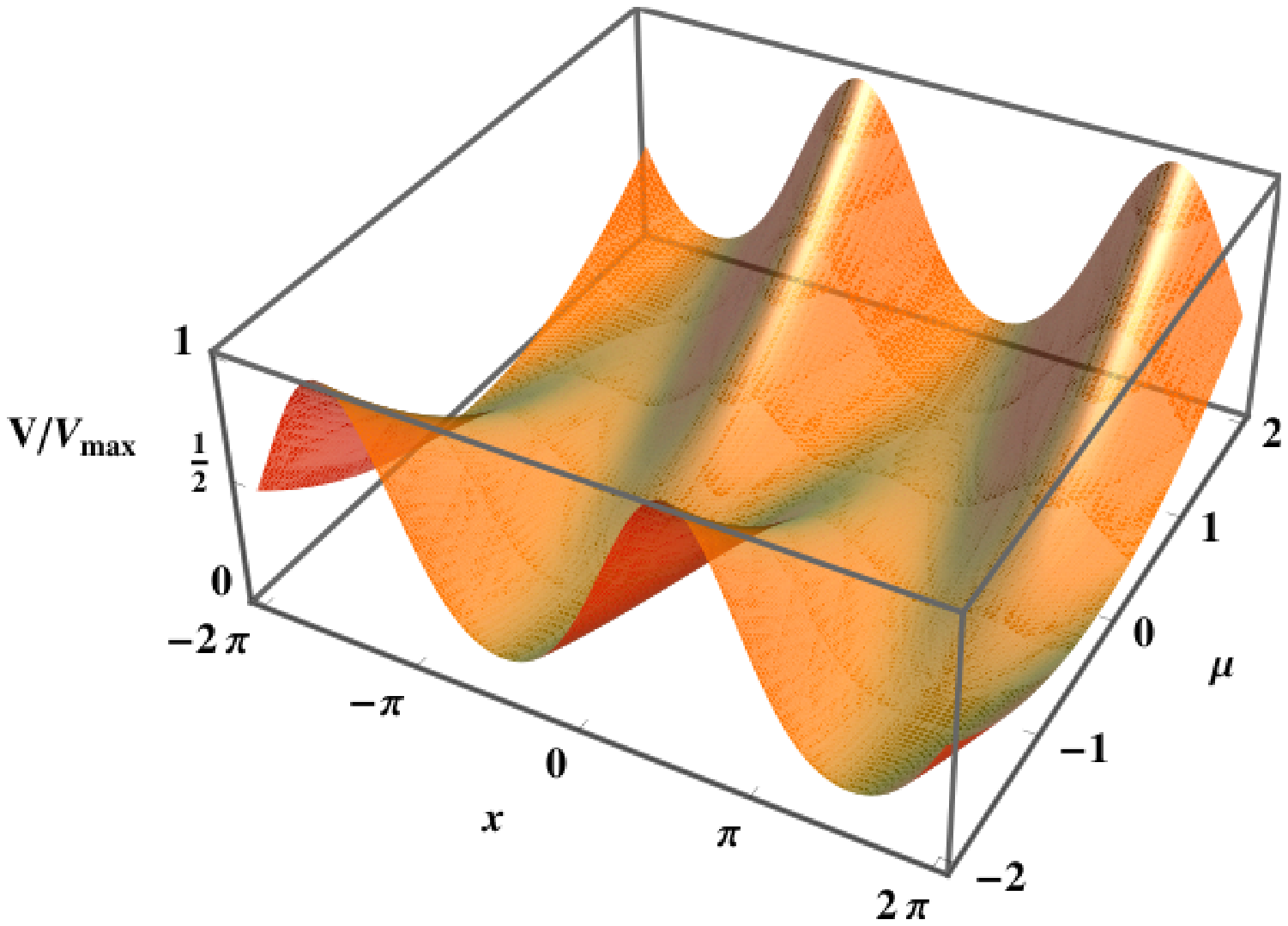}
\includegraphics[width=10.0cm]{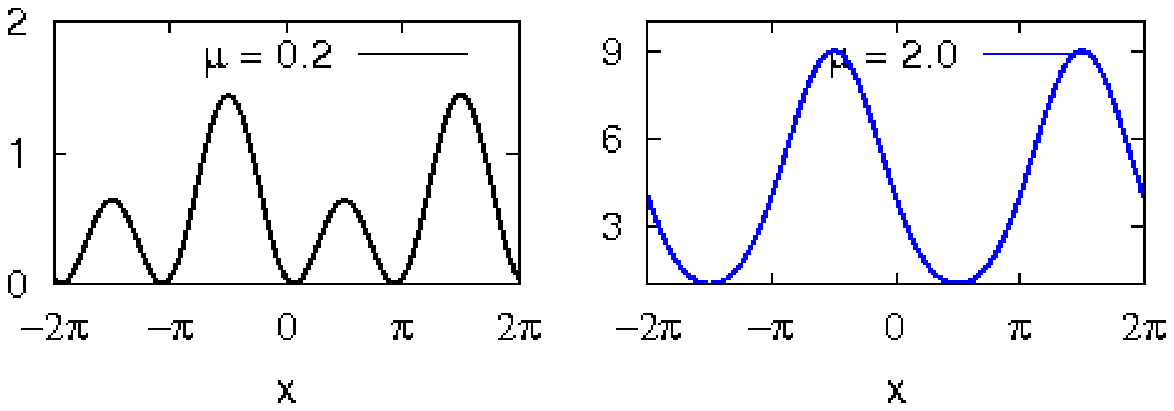}
\caption{ The potentials as a function of the dimensionless pseudo-flux variable $x= \phi/2\phi_0$. On the left, the Li Chen potential $V(x) = V_0 \left(\sin(x-\mu) \right)^2$. The right shows the new potential proposed in this work, $V(x) = V_0 (\sin(x)-\mu)^2$. The left-bottom plot displays the new potential for $\mu = 0.2$ (left), showing two close minima. These minima dissapear for $\mu >1$, as shown on the bottom-right for $\mu = 2.0$ (right).}
\label{Fig-pot}
\end{figure}

\section{Study of the quantum eigenstates: Numerical solution}
\label{sec-numer}

The Schr\"odinger for the potential in Eq. (~\ref{V-new}) has not been studied previously, but it is related to the so-called Hill's equation; therefore, we have studied this problem numerically. First, we make a change of variables so that the flux is expressed in terms of $x= \phi/2\phi_0$ ($ 0 < x < 2\pi $), the external flux is given by the parameter $\mu = f/2\phi_0$, and the energies are expressed in terms of the electrostatic energy $\sigma = q_e^2/C$, so that $\varepsilon = 4 E/\sigma$, then the Schr\"odinger equation becomes

\begin{equation}
\label{pot-new}
\left[-\frac12 \frac{d^2}{dx^2} + v(x) \right] \Psi(x) = \varepsilon \Psi(x)
\end{equation}

In this manner, the solutions depend on a single circuit parameter, namely $G/G_L$, and a single external parameter $\mu$, so that the external field $ f = 2 \phi_0 \mu $. We note that the potential energy term is periodic, with period $2\pi$, and that we look for solutions with the same period (phase factor equal to unity), in order to preserve the charge quantization condition. As indicated previously, we believe it interesting to study two cases: first, we consider the numerical results obtained for the Li-Chen potential (Eq.~\ref{V-LiChen}), and next, compare with the numerical results obtained using our potential (Eq.~\ref{V-new}). For convenience, define the parameter $\theta = 2 (G/G_L)^2/\pi^2$ to represent the circuit parameters so that, after changing coordinates ($ x = \phi/2\phi_0$) and rescaling, the potentials become

\begin{eqnarray}
\label{Eq-pot}
v_{LiChen}(x,\mu) &=&  \theta \sin^2 (x-\mu) \\
v_{new} (x,\mu) & = & \theta \left( \sin(x)- \mu  \right)^2
\end{eqnarray}

The numerical procedure has been described in ~\cite{UTRERAS3}, in which we define a {\em characteristic} function $W(\varepsilon)$ such that $W(\varepsilon ) = 0$ gives the eigenvalues. Notice also that, for the numerical work with the new potential, it is convenient to work with the shifted potential $\tilde v_{new}$,
\begin{eqnarray}
\tilde v_{new} (x,\mu)& =&  -\theta \left( \frac12 \cos(2 x) + \mu \sin(x) \right) \nonumber \\
 & = & v_{new} (x,\mu) - \theta (\mu^2 + \frac12 ).
\end{eqnarray}
in this way, letting the shifted energies $\tilde \varepsilon$ correspond to the shifted potential, the (non shifted) energies will be $\varepsilon = \tilde \varepsilon + \theta (\mu^2 + \frac12 )$.

\subsection*{Li-Chen potential}
\begin{figure}[h]
\includegraphics[width=8.0cm]{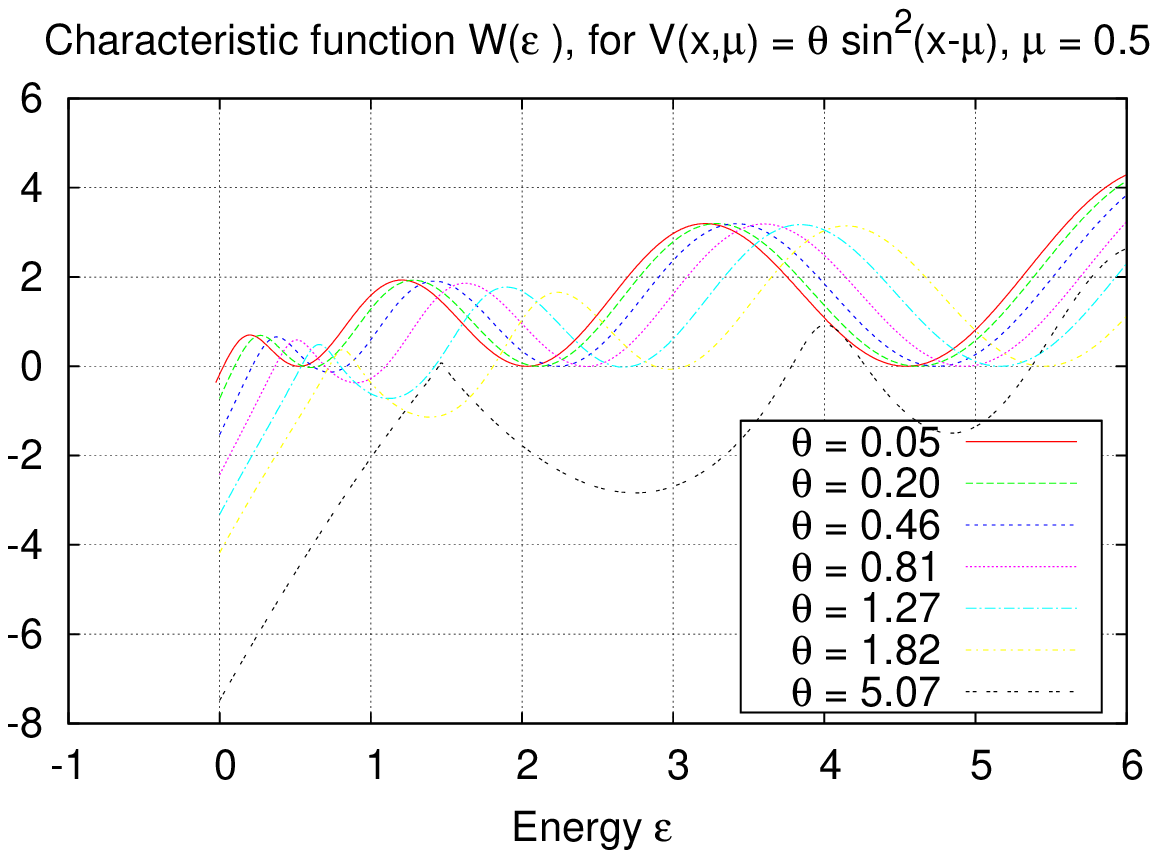}%
\includegraphics[width=8.0cm]{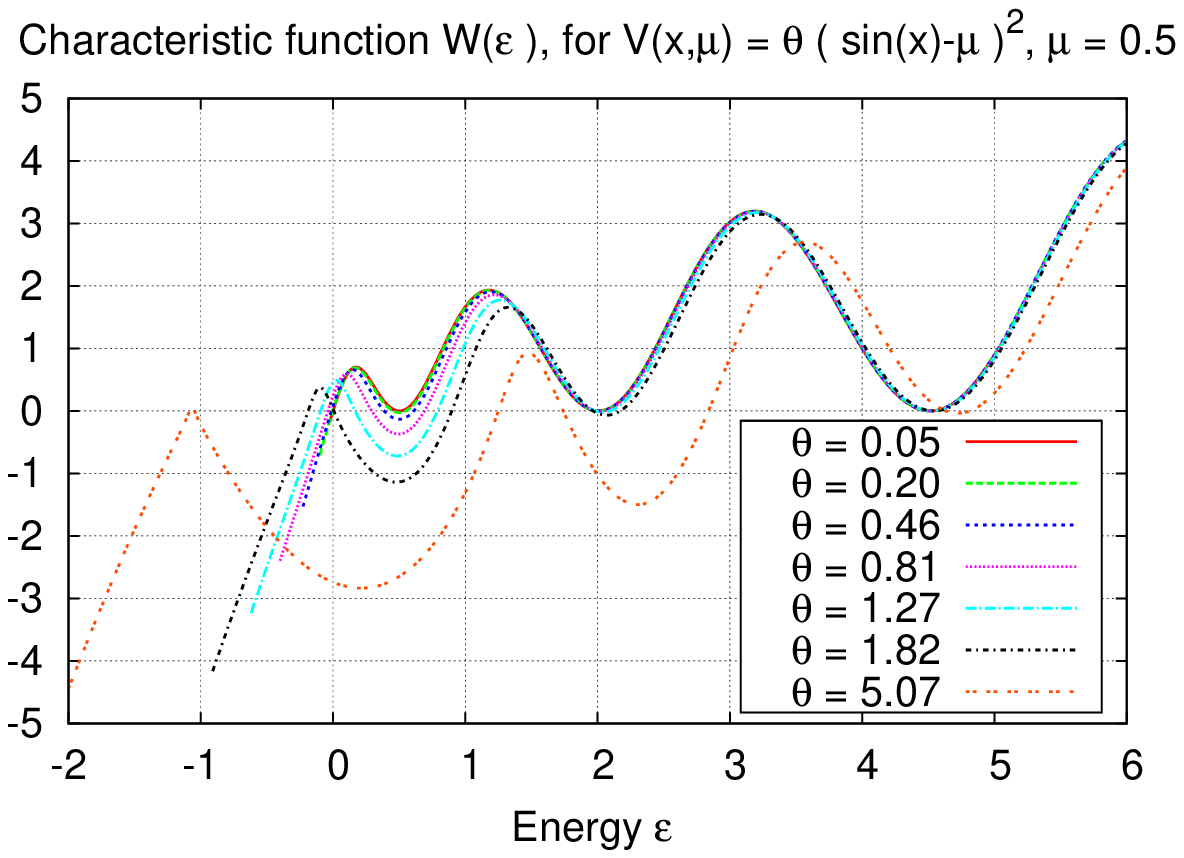}
  \caption{Characteristic function $W(\varepsilon)$ for different values of the parameter $\theta$. On the left hand side, our results for the Li-Chen potential, and for the new potential on the right hand side.}
  \label{Fig-caract}
\end{figure}

In Figure~\ref{Fig-caract}, we show the characteristic function $W(\varepsilon)$ vs. the energy, $\varepsilon$, for the Li-Chen potential. We have studied the behaviour of the $W(\varepsilon)$ for different values of $\theta$. We observe first that, for fixed $\mu$ and small $\theta$, the lower eigenstates values are closely spaced in energy, as a function of $\varepsilon$. When $\theta$ increases, while keeping $\mu$ fixed, these two close eigenvalues converge into one, and eventually, they become doubly degenerate. The same phenomena  happens for the next two (higher energy) eigenvalues, which coalesce at a higher value of $\theta$, and so on for the next two.

\begin{figure}[t]
\includegraphics[width=8.0cm]{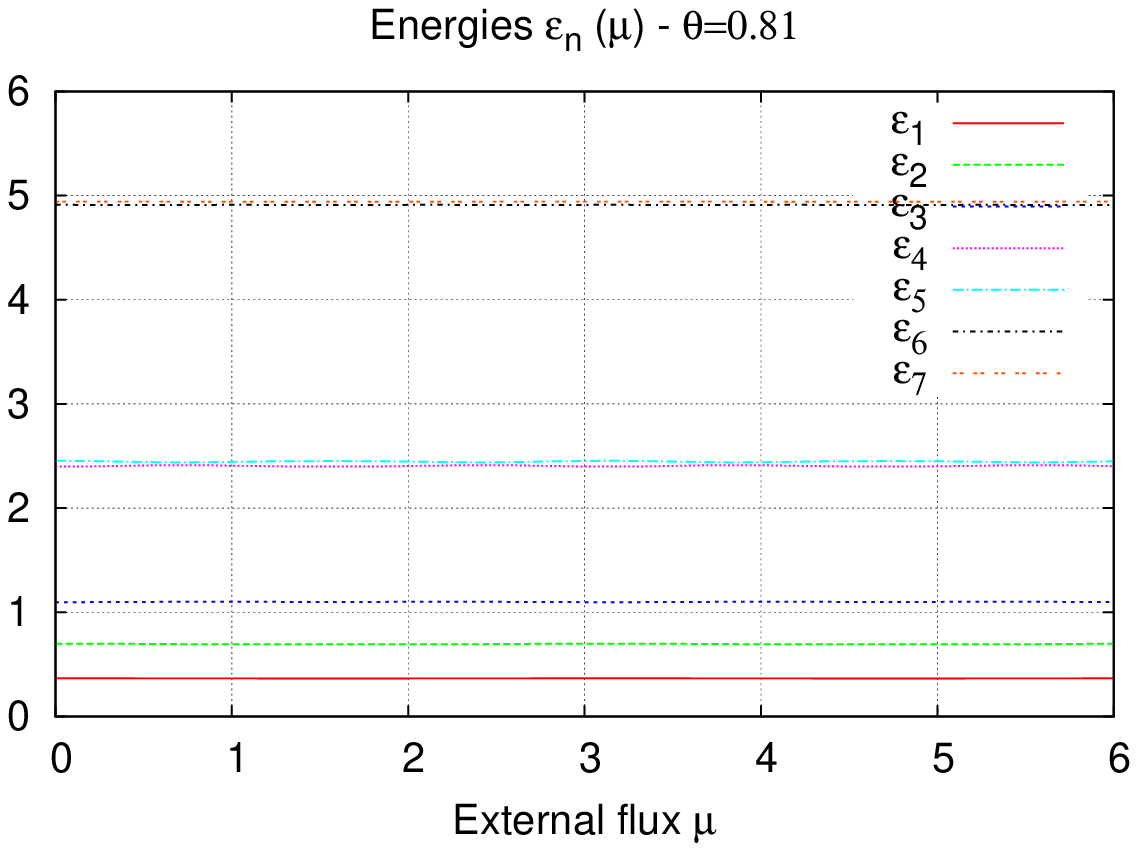}%
\includegraphics[width=8.0cm]{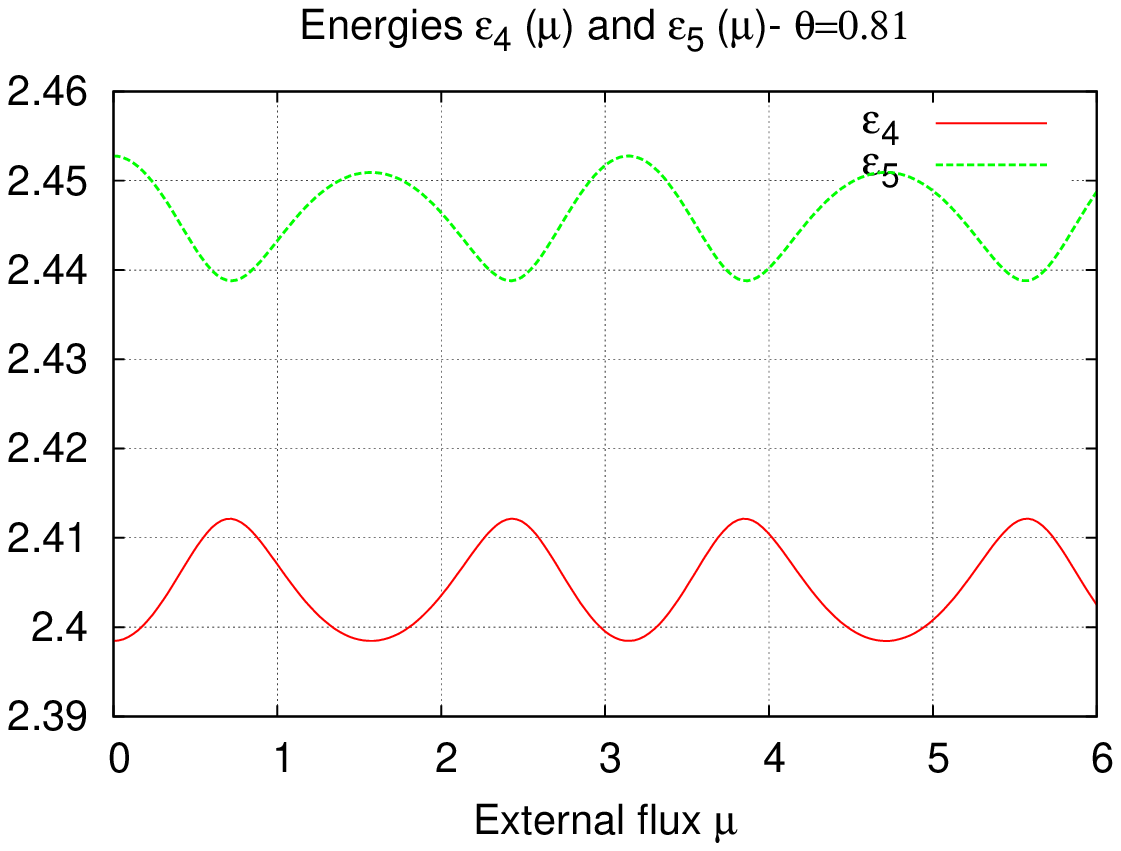}
  \caption{Left hand side, first energy levels as a function of the external flux $\mu$, for different values of the parameter $\theta$, for the Li-Chen potential. On the right hand side, detail of some energy levels as a function of $\mu$. These energy {\em bands} are periodic functions of $\mu$, with period $\pi$, for the Li-Chen case.}
  \label{Fig-Emu-LC}
\end{figure}

Next, in Figure ~\ref{Fig-Emu-LC}, we show the first seven energy {\em bands}, as a function of the external flux $\mu$. At a first glance, they appear to be completely independent of $\mu$; however, upon closer examination, we see that they actually have a periodic structure, as it is shown on the figure on the right. This fact is easily shown, since the potential is periodic in both $x$ and $\mu$, with period $\pi$; one has

\begin{equation}
\label{mu-periodic}
v_{LiChen}(x,\mu) =  \theta \sin^2(x-\mu) = v_{LiChen}(x+ \pi,\mu) = v_{LiChen}(x,\mu + \pi)  .
\end{equation}
If $\Psi(x,\mu) $ is an eigenstate of $H(x,\mu)$, with energy eigenvalue $\varepsilon (\mu)$, then
\begin{equation}
\label{mu}
H(x,\mu) \Psi(x\,\mu ) = \varepsilon (\mu) \Psi(x,\mu),
\end{equation}
and $\Phi(x,\mu^{\prime})$ corresponding to the eigenvalue $\mu^{\prime}$,
\begin{equation}
\label{muprime}
H(x,\mu^{\prime}) \Phi(x,\mu^{\prime}) = \varepsilon (\mu^{\prime}) \Phi(x,\mu^{\prime}).
\end{equation}
On account of Eq.(\ref{mu-periodic}), with $\mu^{\prime} = \mu + \pi$, we have $H(x,\mu +\pi ) = H(x,\mu)$, and Eq.(\ref{muprime}) then
\begin{eqnarray}
H(x,\mu ) \Phi(x,\mu+ \pi) & = & \varepsilon (\mu + \pi ) \Phi(x,\mu + \pi ) ,
\end{eqnarray}
therefore, we may identify $\Psi(x,\mu) = \Phi(x,\mu + \pi)$, and then
\begin{equation}
\varepsilon (\mu) = \varepsilon (\mu + \pi).
\end{equation}

\newpage

\subsection*{New potential}
The situation for the new (Eq.~\ref{V-new}) potential is shown in Figure~\ref{Fig-caract}. First, we show the {\em characteristic} function $W(\varepsilon)$, for fixed $\mu$, plotted as a function or $\varepsilon$. We observe that, as we increase the value of $\theta$, the eigenvalues colasesce in pairs, becoming doubly degenerate, just as in the case of the Li-Chen potential. In Figure ~\ref{Fig-caract} we show the magnetic energy bands, $\varepsilon_n(\mu)$, for increasing values of the circuit parameter $\theta$. We observe that, as expected, the bands are not periodic.

We have observed numerically that the energy bands $\varepsilon(\mu)$ are symmetric with respect to the change in sign of the external flux, i.e. $\varepsilon(\mu) = \varepsilon(-\mu)$. This property is easily explained, since the potentials (Li-Chen, and  new), satisfy the property
\begin{equation}
v(x,\mu) = v(-x,-\mu).
\end{equation}
Let $\Psi(x,\mu)$ be the eigenfunction corresponding to $\varepsilon(\mu)$,
\begin{equation}
H(x,\mu) \Psi(x,\mu) = \varepsilon (\mu) \Psi(x,\mu),
\end{equation}
and let $\tilde T$ the inversion operator, defined so that $\tilde T \left[ \Psi(x) \right] = \Psi(-x)$, then
\begin{eqnarray}
\tilde T \left[ H (x,\mu) \Psi(x,\mu)\right] &=& \varepsilon (\mu) \tilde T \Psi(x,\mu) \\
 H(-x,\mu) \Psi(-x,\mu) &=& \varepsilon (\mu) \Psi(-x,\mu) \\
 H(x,-\mu) \Psi(-x,\mu) &=& \varepsilon (\mu) \Psi(-x,\mu) \label{eq-last},
\end{eqnarray}
since $H(-x,\mu) = H(x,-\mu)$ and $v(-x,\mu) = v(x,-\mu)$. Let us define the wavefunction $\Phi(x,-\mu) \equiv \Psi(-x,\mu) $, then $\Phi(x,-\mu)$  is an eigenfunction of $H(x,-\mu)$  corresponding to the eigenvalue $\varepsilon(-\mu)$,
\begin{equation}
H (x,-\mu)\Phi(x,-\mu) = \varepsilon(-\mu) \Phi(x,-\mu)
\end{equation}
and the last equation~(Eq. \ref{eq-last} )tells us that $H(x,-\mu) \Phi(x,-\mu) = \varepsilon (\mu) \Phi(x,-\mu) $, or
\begin{equation}
\varepsilon (-\mu) = \varepsilon(\mu),
\end{equation}
finally, we may identify
\begin{equation}
\Psi(x,-\mu) = \Psi(-x,\mu).
\end{equation}

\begin{figure}[h]
\includegraphics[width=7.0cm]{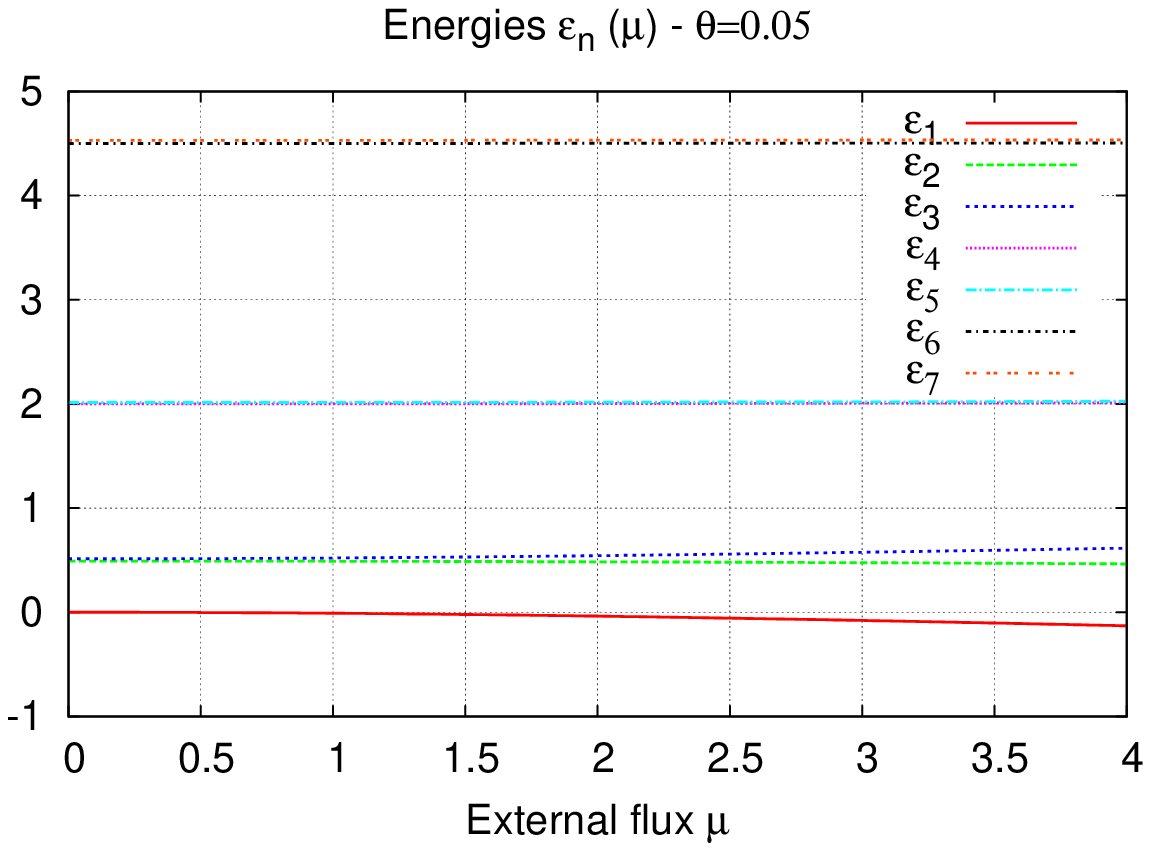}%
\includegraphics[width=7.0cm]{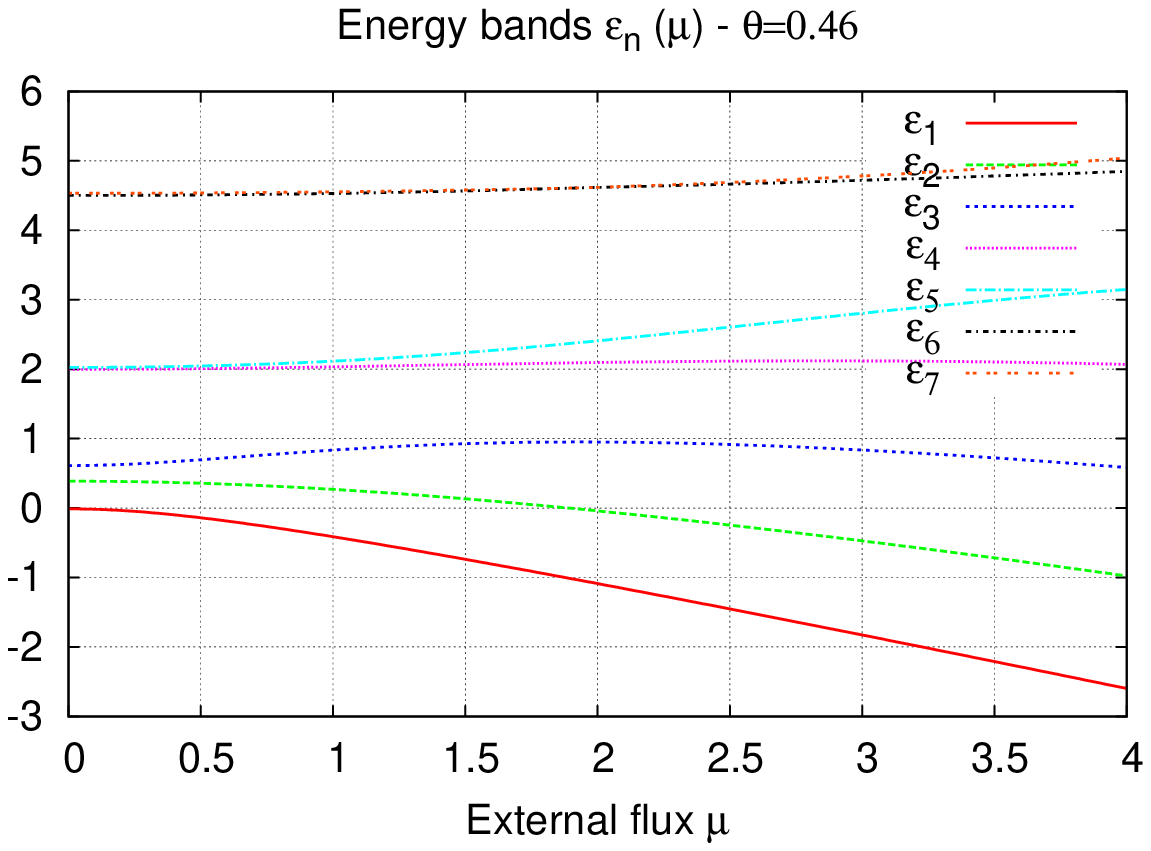}\\
\includegraphics[width=7.0cm]{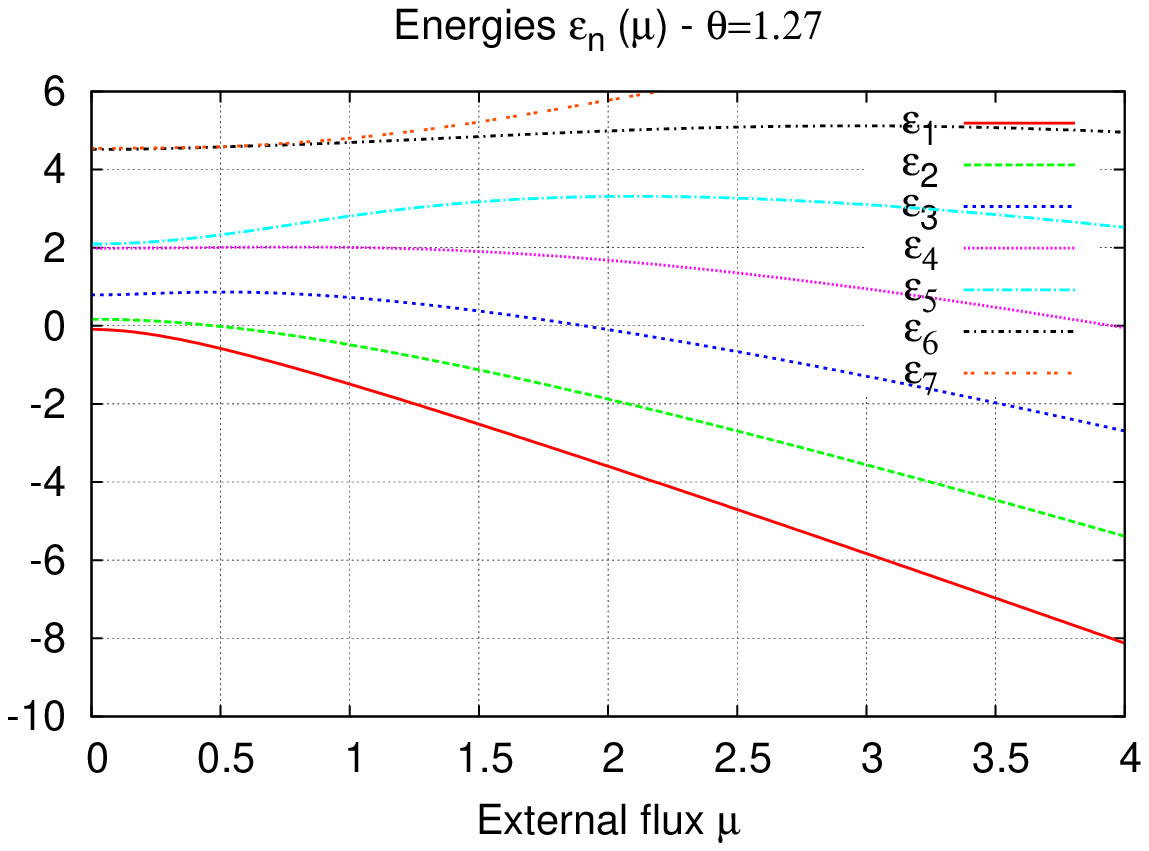}%
\includegraphics[width=7.0cm]{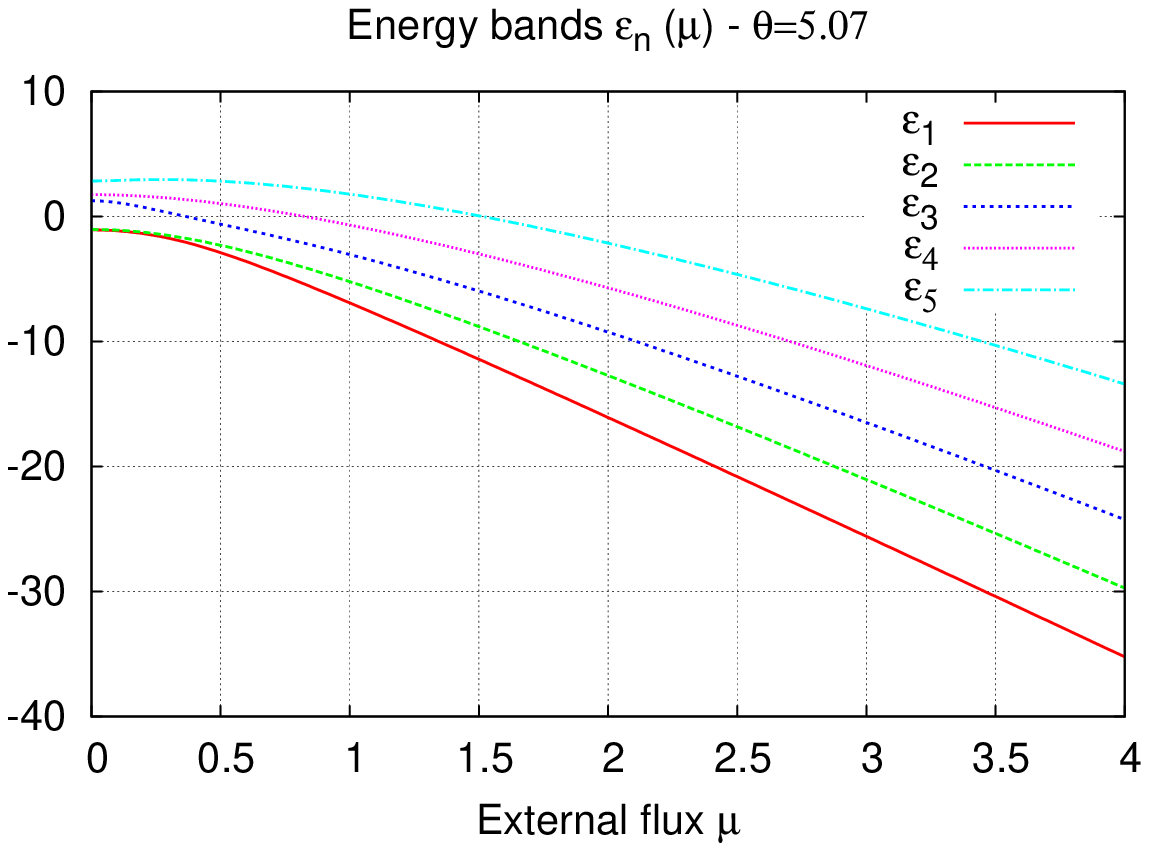}
  \caption{New potential. Energy bands as a function of the external flux $\mu$. The figures differ in the value of the parameter $\theta$.}
  \label{Fig-Emu-UL}
\end{figure}

To understand the behaviour of the energies as a function of $\mu$, that we have described in the previous paragraphs, let us observe it for some representative values, one small $\mu = 0.2$, and one 'large', $\mu = 2.0$, as shown in Figure (\ref{Fig-pot}). It is clear then that, since
\begin{equation}
v_{new} (x,\mu )  = \theta \left( \sin^2(x) - 2 \mu \sin(x) + \mu^2 \right) ,
\end{equation}
then,  the dominant behaviour observed is due to the $\mu \sin(x)$ term, which has a minimum at $x = \pi/2$. In the neighborhood of $x = \pi/2$, the potential has a harmonic oscillator form,
\begin{equation}
v_{new} (x,\mu )  = \theta \left(   \left( \mu^2-2 \mu +1 \right) + \left( \mu-1\right) {\left( x-\frac{\pi }{2} \right) }^{2}+...\right) ,
\end{equation}
therefore, the lowest lying energies become ( for $|\mu| >> 1$)
\begin{equation}
\varepsilon_n (\mu) \approx \sqrt{2\theta (\mu - 1) } \, ( n + 1/2) + \theta \left( \mu^2-2 \mu +1 \right).
\end{equation}
For a direct comparison with the numerical values, we need to obtain
\begin{equation}
\tilde \varepsilon_n (\mu) \approx \varepsilon_n (\mu) - \theta (\mu^2 + 1/2) = \sqrt{2\theta (\mu - 1) } \, ( n + 1/2) + \theta \left( - 2 \mu +1/2 \right).
\end{equation}
These eigenvalues are a good approximations for sufficiently large $\mu > 1$, since then the potential gets larger, and the barrier between neighboring minima becomes larger too. At the same time, as $\theta $ becomes large, the approximation also improves, since this also increases the barrier. For negative $\mu$, the position of the minima changes, but the results for the energies still hold, with $\mu$ replaced by $|\mu|$. From the formula above, we see that the difference between  two neighboring energy levels, say $ n $, and $n+1 $, is independent of $n$,
\begin{equation}
\tilde \varepsilon_{n+1} - \tilde \varepsilon_n \approx \sqrt{2\theta (|\mu| - 1) },
\end{equation}
a relation that is verified from our numerical results. On the other hand, for $|\mu| << 1$ we have
\begin{equation}
\varepsilon_n (\mu) \approx \sqrt{2\theta (1 - \mu^2 ) } \, ( n + 1/2) + \theta \left( \mu^2-2 \mu +1 \right).
\end{equation}


\section{Final comments}
In this work, we have proposed a new way to include the effect of a static magnetic field on a quantum circuit. This gives us two quite different potentials to describe the effect of a magnetic field. We have made a comparative study of the behaviour of the energies as a funtion of the external magnetic field. The results show that this behaviour is very different, and therefore, one could design experiments to test the applicability of either potential to explain the experimental results. In a forthcoming paper we intend to test the predictions of our model to some real physical systems.

\section{Acknowledgements}
This research received financial support from DIDUACH Grant \# S-2008-57, Anillo ACT 15 and ACT 24 of Bicentennial Program of Sciences and Technology-Chile, Millenium Scientific Initiative under contract Millennium Science Nucleus P06-022-F and \emph{Convenio de Desempe\~no} UTA MECESUP-2 and FONDECYT 11080229.

\appendix

\end{document}